\documentclass[pra,showpacs,twocolumn]{revtex4}
\usepackage{bm,graphicx,amsmath}
\usepackage{psfrag}
\begin{document}

\title{Dark solitons near the Mott-insulator--superfluid phase transition}

\author{Konstantin V.~Krutitsky$^{1}$, Jonas Larson$^{2,3}$, and Maciej Lewenstein$^{4,5}$}

\affiliation{$^1$Fakult\"at f\"ur Physik der Universit\"at Duisburg-Essen,
             Campus Duisburg, Lotharstra{\ss}e 1, 47048 Duisburg, Germany}

\affiliation{$^2$NORDITA, 106 91 Stockholm, Sweden}

\affiliation{$^3$Department of Physics, Stockholm University, AlbaNova University Center, 106 91 Stockholm, Sweden}

\affiliation{$^4$ICFO-Institut de  Ci\`encies de Fot\`oniques,
008860 Castelldefels (Barcelona), Spain}

\affiliation{$^5$ICREA-Instituci\'o Catalana de Recerca i Estudis
Avan\c{c}ats, Llu\'{i}s Companys 23, 08010 Barcelona, Spain}
%\email{kostya@theo-phys.uni-essen.de}
\date{\today}

\begin{abstract}
Dark solitons of ultracold bosons in the vicinity of the
Mott-insulator--superfluid phase transition are studied. Making use
of the Gutzwiller ansatz we have found antisymmetric eigenstates
corresponding to standing solitons, as well as propagating solitons
created by phase imprinting.
Near the phase boundary, superfluidity has either a particle or a hole character
depending on the system parameters, which greatly affects the characteristics
of both types of solitons. Within the insulating Mott regions, soliton solutions are prohibited by lack of phase coherence between the lattice sites.
Linear and modulational stability show that the soliton solutions are sensitive to small perturbations and, therefore, unstable.
In general, their lifetimes differ for on-site and off-site modes. For the on-site modes, there are small areas between the Mott-insulator regions
where the lifetime is very large, and in particular much larger than that for the off-site modes.
\end{abstract}


\maketitle

%----------------------------------------
\section{Introduction}
%----------------------------------------

Ultracold atomic gases provide a perfect
playground to study nonlinear atom optics, and nonlinear structures
and textures, such as solitons~\cite{stringari,kivshar}. These
studies have led to the observations of
dark~\cite{maciek-sengstock,denschlag,Stellmer08,Weller08,Theocharis} and
bright~\cite{khaykovich/strecker} solitons in trapped Bose-Einstein
condensates, bright solitons stabilized by the
presence of dark ones~\cite{sengstock} as well as
oscillating soliton/vortex rings~\cite{Shomroni09}. The analogy to nonlinear
optics~\cite{kivshar} has triggered theoretical interest in discrete
(lattice) solitons~\cite{smerzi,anna}, and has led to the seminal
observations of gap solitons, i.e., lattice solitons with repulsive
interactions, but with an appropriate dispersion
management~\cite{oberthaler}.

While most of the studies of solitons were concentrated on their
classical aspects, more recently, considerable interest has been
devoted to the effect of thermal noise~\cite{muryshev,anna}, quantum
properties of solitons, and the role of quantum fluctuations~\cite{MR}.
The latter may cause filling up of the dark soliton core in the quantum
detection process, as was shown using the Bogoliubov-de Gennes
equations~\cite{dziarmaga}. The same method was also employed to
study the stability of solitons~\cite{yulin,JK99,KCTFM}, excitations caused by
the trap opening~\cite{castin}, and entanglement generation in
collisions of two bright solitons~\cite{boris}. A noisy version of
the standing bright solitons was studied using the exact
diagonalization and quantum Monte Carlo method~\cite{juha}. Bright
solitons in 1D were considered in Ref.~\cite{castin}, where exact
Lieb-Liniger solutions were used to calculate the internal
correlation function of the particles positions. Making use of the
discrete nonlinear Schr\"odinger equation (DNSE), and the
time-evolving block decimation algorithm~\cite{vidal} it was
demonstrated that quantum effects cause the soliton to fill in, and
that soliton collisions become inelastic~\cite{carr}.

All previous studies of lattice solitons were done in the
deep superfluid phase. However, near the phase boundary between
superfluid (SF) and Mott insulator (MI)~\cite{Fisher}, the
propagation of matter waves becomes strongly suppressed due to the
enhancement of quantum fluctuations. In addition, near the phase
transition the superfluidity is determined either by particles, or
holes depending on the lattice filling and system
parameters~\cite{WCHZ}. Therefore, fundamental questions arise: Do
the solitons of any form exist in this regime, and, if the answer
is yes, what are their properties? Evidently, the characteristics
of such solitons are expected to be very different from the ones
in the deep SF regime. The aim of the present work is to study
lattice solitons near the MI-SF phase transition. We find that, in
particular, no soliton solutions exist within the MI regions, and
that both standing and propagating solitons exhibit anomalous
behavior in the hole SF region. A generalized Bogoliubov-de~Gennes stability analysis
indicates that the solitons are sensitive to perturbations and
thereby break down over longer time periods. Our results are
achieved by employing the position-dependent Gutzwiller ansatz
which gives a satisfactory description of the quantum phases of
inhomogeneous bosonic systems~\cite{JBCGZ,BPVB07} as well as their
dynamical behavior~\cite{Z}. We note that the same method was
recently utilized to study vortices in the vicinity of the MI-SF
phase transition~\cite{WCHZ,GM}.

The paper is organized as follows. In Sec. II we present the
model and the Gutzwiller method. In Sec. III we discuss
properties of the ground states of the model, obtained using the
Gutzwiller method, and the nature of the SF phase (hole
SF versus particle SF) is discussed. Section IV is devoted to the numerical
studies of the standing dark solitons (kinks), i.e.
stationary anti-symmetric solutions of the time-dependent
Gutzwiller equations. We analyze in detail their shape and nature
in different regions of the phase diagram. In Sec. V we consider
the linear and modulational stabilities of the dark solitons, and
show that generically they are unstable, but may have quite long
lifetimes. The Sec. VI concerns to the problem of
experimental generation of propagating solitons using phase
imprinting method. We identify here the regimes were such
generation, and the well defined soliton propagation is  possible.
We conclude in Sec. VII.

%----------------------------------------
\section{The model}
%----------------------------------------

We consider a system of ultracold interacting bosons in a
$d$-dimensional lattice described by the Bose-Hubbard Hamiltonian
\begin{eqnarray}
\label{BHH}
\hat H
&=&
-J
\sum_{\bf i}
\sum_{\alpha=1}^d
\left(
    \hat a_{\bf i}^\dagger
    \hat a_{{\bf i}+{\bf e}_\alpha}
    +
    {\rm h.c.}
\right)
\nonumber\\
&+&
\frac{U}{2}
\sum_{\bf i}
\hat a^\dagger_{\bf i}
\hat a^\dagger_{\bf i}
\hat a_{\bf i}
\hat a_{\bf i}
-
\mu
\sum_{\bf i}
\hat a^\dagger_{\bf i} \hat a_{\bf i}
\;,
\end{eqnarray}
where ${\bf e}_\alpha$ is a unit vector on the lattice in the direction $\alpha$,
$J$ is the tunneling matrix element,
$U$ is the on-site atom-atom interaction energy,
and $\mu$ the chemical potential.
The annihilation and creation operators at site ${\bf i}$,
$\hat a_{\bf i}$ and $\hat a^\dagger_{\bf i}$,
obey the bosonic commutation relations.
Throughout the paper, we will be dealing with repulsive interaction, i.e., $U>0$.

Our analysis employs the Gutzwiller ansatz. Thereby, eigenstates of
the Hamiltonian~(\ref{BHH}) are taken as products of local states
\begin{equation}
|\Phi\rangle
=
\prod_{\bf i}
|s_{\bf i}\rangle
\;,\quad
|s_{\bf i}\rangle
=
\sum_{n=0}^\infty
c_{{\bf i}n}
|n\rangle_{\bf i}
\end{equation}
satisfying the normalization conditions
\begin{equation}
\label{norm}
\sum_{n=0}^\infty
\left|
    c_{{\bf i}n}
\right|^2
=1
\;.
\end{equation}
Here, $|n\rangle_{\bf i}$ is the Fock state with $n$ atoms at site ${\bf i}$.
The corresponding energy functional takes the form
\begin{eqnarray}
\label{E}
E
&=&
-J
\sum_{\bf i}
\sum_{\alpha=1}^d
\left(
    \psi_{\bf i}^*
    \psi_{{\bf i}+{\bf e}_\alpha}
    +
    {\rm c.c.}
\right)
\nonumber\\
&+&
\frac{U}{2}
\sum_{\bf i}
\langle
\hat a^\dagger_{\bf i}
\hat a^\dagger_{\bf i}
\hat a_{\bf i}
\hat a_{\bf i}
\rangle
-
\mu
\sum_{\bf i}
\langle
\hat a^\dagger_{\bf i} \hat a_{\bf i}
\rangle
\;,
\end{eqnarray}
where
\begin{equation}
\label{psi}
\psi_{\bf i}
=
\langle \hat a_{\bf i}\rangle
=
\sum_{n=1}^\infty
c_{{\bf i},n-1}^* c_{{\bf i}n}
\sqrt{n}
\end{equation}
is the condensate order parameter.
The mean number of condensed atoms in this model is given by $|\psi_{\bf i}|^2$
which cannot be larger than the mean occupation number
$\langle\hat n_{\bf i}\rangle$~\cite{phas} given by
\begin{equation}
\label{n}
\langle\hat n_{\bf i}\rangle
=
\sum_{n=1}^\infty
\left|
    c_{{\bf i}n}
\right|^2
n
\;.
\end{equation}
Equations describing the time-dependent Guztwiller ansatz,
are then easily obtained by  minimization of the energy functional
(\ref{E}) with the constraint (\ref{norm}), and replacing the
contribution of the latter by the time derivative. Such approach
is equivalent to the use of time-dependent variational principle
applied to an appropriately defined Lagrange action, as described
for instance in Ref. \cite{Perez-Garcia}. It leads to the
following equations of motion~\cite{Z,BPVB07}:
\begin{eqnarray}
\label{GEd}
i\hbar
\frac{d c_{{\bf i}n}}{dt}
=
&-&
J
\left(
    \Psi_{\bf i} \sqrt{n+1} c_{{\bf i},n+1}
    +
    \Psi_{\bf i}^* \sqrt{n} c_{{\bf i},n-1}
\right)
\nonumber\\
&+&
\left[
    \frac{U}{2}n(n-1)
    -
    \mu n
\right]
c_{{\bf i}n}
\;,
\end{eqnarray}
where
$
\Psi_{\bf i}
=
\sum_{\alpha=1}^d
\left(
    \psi_{{\bf i}+{\bf e}_\alpha}
    +
    \psi_{{\bf i}-{\bf e}_\alpha}
\right)
$.
Note that Eqs.~(\ref{GEd}) are invariant under transformation
$c_{{\bf i}n}\to (-1)^n c_{{\bf i}n}$.

%----------------------------------------
\section{\label{GS}Ground state}
%----------------------------------------

\begin{figure}[t]
%\centering

%\psfrag{n}[c]{$n$}
%\psfrag{c}[r]{$\left|c_n^{(0)}\right|^2$}

%\includegraphics[width=7cm]{cn.eps}

\hspace{-2.5cm}
\includegraphics[width=9cm]{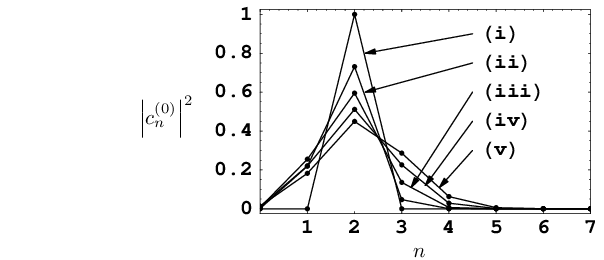}

\caption{Ground-state solutions for the atomic distribution
$\left|c_n^{(0)}\right|^2$.
The scaled chemical potential $\mu/U=1.2$
and the tunneling rates $2dJ/U$:
0.7~(i), 0.5~(ii), 0.3~(iii), 0.15~(iv), and 0.05~(v).
The lines connecting the dots are to guide the eye.}
\label{cn}
\end{figure}

In the ground state, the coefficients $c_{{\bf i}n}$ do not depend on the site index ${\bf i}$.
According to Eq.~(\ref{psi}), $\psi_{\bf i}\equiv\psi$ and, therefore, in Eq.~(\ref{GEd})
$\Psi_{\bf i}=2d\psi$.
The ground-state solution has the form
\begin{equation}
c_{{\bf i}n}(t)
=
c_n^{(0)}
\exp
\left(
    -i \omega_0 t
\right),
\end{equation}
and the coefficients $c_n^{(0)}$ can be calculated numerically
using different methods. Probably the most efficient one is to
solve the single-site eigenvalue problem for the mean-field
Hamiltonian corresponding to the Gutzwiller ansatz. This has been
done also by us by means of exact diagonalization, in the same
manner as in Refs.~\cite{SKPR,OSS}. The results of the
calculations are shown in Fig.~\ref{cn}. The coefficients
$|c_n^{(0)}|^2$ form a broad Poissonian-like distribution in the
SF phase, where $\psi^{(0)}\ne 0$. In the MI phase,
however, $c_n^{(0)}=\delta_{n,n_0}$ resulting in $\psi^{(0)}=0$.
In the mean-field approach, the boundaries between the MI and SF
are determined by
\begin{displaymath}
2dJ_c/U
=
\frac
{(n_0-\mu/U)(\mu/U-n_0+1)}
{1+\mu/U}
\;,
\end{displaymath}
where $n_0$ is the smallest integer greater than $\mu/U$~\cite{Sachdev}.

%----------------------------------------
%----------------------------------------

In the numerical calculations presented in this section and later on,
$n$ was restricted by some finite $N$ ($c_{n} \equiv 0$ for $n>N$).
The cut-off number of atoms $N$ was chosen large enough such that
its influence on the eigenstates is negligible.
For example, for the plots shown in Fig.~\ref{cn}, it was enough to use $N=10$.

We have also checked that the same results for $c_n^{(0)}$
can be obtained propagating Eq.~(\ref{GEd})
in the imaginary time~\cite{Dalfovo} starting with the initial condition
$c_n(0)$ which gives nonvanishing $\psi(0)$.
In our calculations, we used $c_n(0)=1/\sqrt{N+1}$ but other choices
of $c_n(0)$ would lead to the same results as long as the ansatz and the ground state
has a non-zero overlap.
In the calculations of the ground state of a homogeneous lattice,
the imaginary-time propagation technique is less efficient than the exact
diagonalization. However, it becomes more efficient in the calculations
of the states with coefficients $c_{{\bf i}n}$ depending on the site index ${\bf i}$.

\begin{figure}[t]
\begin{center}
\includegraphics[width=8cm]{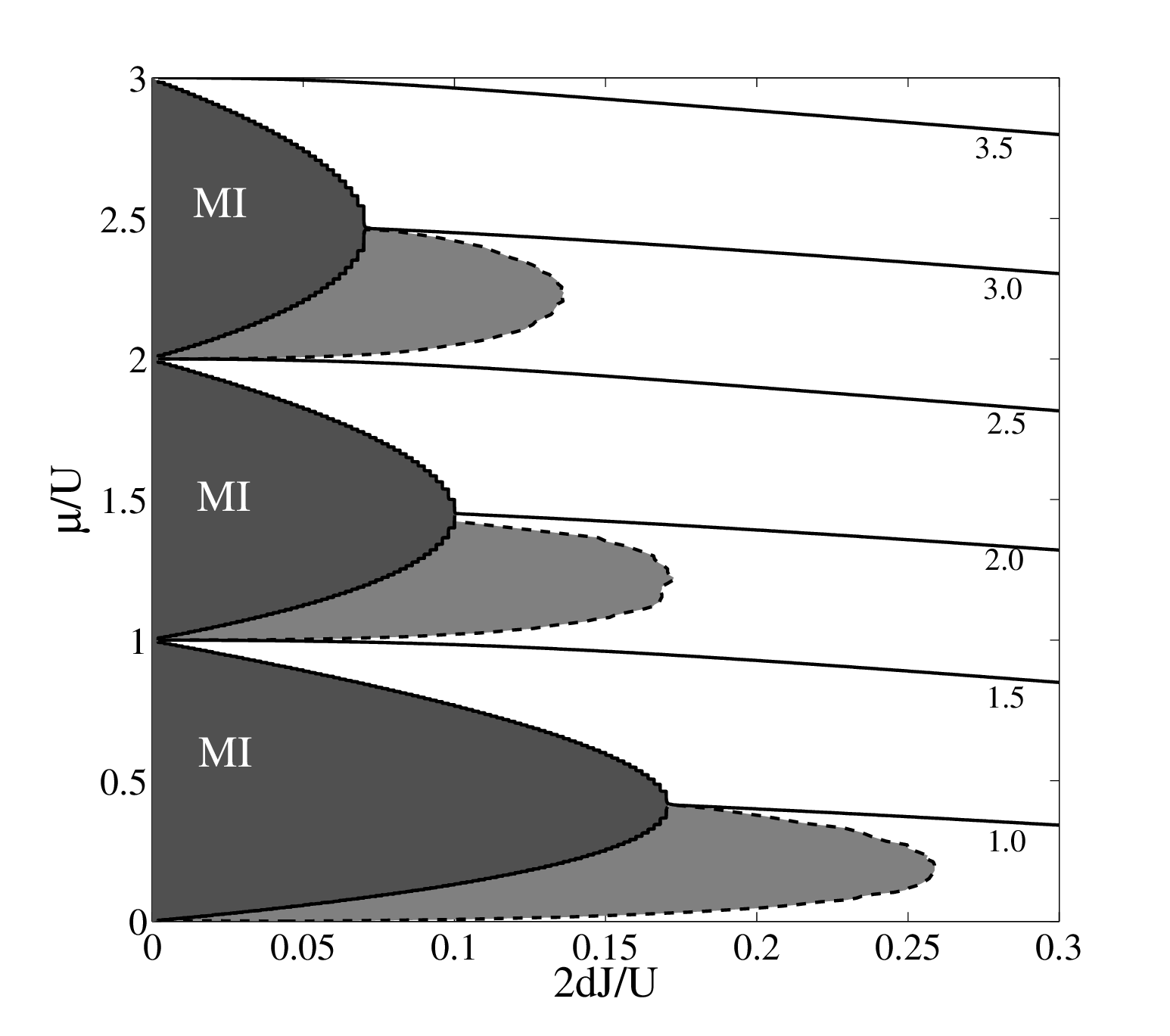}
\caption{
Dark areas bounded by the solid lines show the first three MI zones ($n_0=1,2,3$).
The lines of constant $\langle\hat{n}\rangle$ are labeled
by the corresponding atomic densities.
In the grey areas, where $\mu'(J)>0$ at constant $\langle\hat{n}\rangle$
the superfluidity has a hole character.
In the rest part of the diagram, we have a particle SF.
}
\label{hs}
\end{center}
\end{figure}

As discussed in Ref.~\cite{WCHZ}, near the phase boundary one has to distinguish
between particle and hole superfluidity. For the hole SF, the function
$\mu(J)$ at constant filling factor $\langle\hat n\rangle$ has a positive derivative
$\mu'(J)$. This is only possible for fillings $n_0-0.5 < \langle\hat n\rangle < n_0$ as is demonstrated in Fig.~\ref{hs} showing the corresponding
hole SF regions. For the particle SF, on the other hand, $\mu'(J)<0$.
Consequently, far away from the phase boundary, superfluidity has always a particle character.
As we will see in the next sections, the difference between particle and hole
superfluidity plays an essential role for the character of the soliton modes.

%----------------------------------------
%----------------------------------------

%----------------------------------------
\section{Standing solitons}
%----------------------------------------

In the present section, we study low-energy excited states,
where the coefficients $c_{{\bf i}n}$ as well as
the order parameters $\psi_{\bf i}$ depend only on one spatial dimension $\alpha$.
Without loss of generality, we can assume that this is $\alpha=1$. Then
\begin{equation}
\psi_{{\bf i}\pm{\bf e}_\alpha}
=
\left\{
    \begin{tabular}{ll}
    $\psi_{i_1\pm 1}$ &, if $\alpha=1$, \\
    $\psi_{i_1}$      &, if $\alpha>1$.
    \end{tabular}
\right.
\end{equation}
Therefore, $\Psi_{\bf i} = \psi_{i_1-1} + \psi_{i_1+1} + 2(d-1)\psi_{i_1}$.
In the following, we shall replace $i_1$ by $l$.

We are interested in the stationary solutions of Eqs.~(\ref{GEd})
\begin{eqnarray}
c_{ln}(t)
&=&
c_{ln}^{(0)}
\exp
\left(
    - i \omega_l t
\right)
\;,
\nonumber\\
\label{omegal}
\hbar\omega_l
&=&
-2J
\left[
    \psi_{l-1}^{(0)}
    +
    \psi_{l+1}^{(0)}
    +
    2(d-1)
    \psi_{l}^{(0)}
\right]
\\
&+&
\sum_{n=0}^\infty
\left[
    \frac{U}{2}n(n-1) - \mu n
\right]
\left|
    c_{ln}^{(0)}
\right|^2
\;,
\nonumber
\end{eqnarray}
where $\psi_l^{(0)}$ is determined by the coefficients $c_{ln}^{(0)}$ according to Eq.~(\ref{psi}).
We require that $\psi_l^{(0)}$
is an anti-symmetric function with respect to the middle point of the lattice $l_0$.
These are the kink states which can be treated as standing dark solitons.
In contrast to the ground state discussed in Sect.~\ref{GS}, all the quantities
which describe the solitons are labeled by the site index $l$.

In general, one has to distinguish between the two cases:
when the middle point $l_0$ is on the lattice site
(on-site modes) and in the middle of two neighboring sites (off-site modes).
The two modes have different energies, and the difference
defines the Peierls-Nabarro barrier~\cite{lattsol1}, which may affect the mobility of solitons.
In the present
work, however, when considering propagating solitons we will be dealing only with situations where the presence of the barrier is not relevant.
Nevertheless, the stability of the on-site and off-site modes can in general be different,
as it will be shown in the next section.

We consider first the SF phase.
The computations of the excited eigenstates in the SF regime are
performed using the imaginary-time propagation technique~\cite{Dalfovo}.
In general, as long as the initial state for the imaginary-time propagation
has the required symmetry,
the propagation will finally give the lowest-energy excited eigenstate with the same symmetry.
The initial state for the imaginary-time propagation in Eq.~(\ref{GEd}) was chosen
to be the ground state with $c_{ln}(0)\equiv c_{n}^{(0)}>0$
for the sites located to the left from the middle point of the lattice.
As it follows from Eq.~(\ref{psi}), $\psi_l(0)$ are positive for these sites.
For the sites located to the right from the middle point of the lattice,
we used $c_{ln}(0) = (-1)^n c_{n}^{(0)}$, which is dictated by the symmetry of
Eqs.~(\ref{GEd}).
Then the $\psi_l$'s take negative values, but their absolute values are the same
as for the sites to the left from the middle point. For the on-site modes, we had in addition $c_{ln}(0)=\delta_{n,n_0}$
for the middle point which gives $\psi_l(0)=0$ for this site.
Thus, $\psi_l$ has the form of a kink state and this symmetry is preserved
during the imaginary-time evolution.
Stationary solutions (\ref{omegal}) of Eqs.~(\ref{GEd})
are worked out for a lattice with the finite number of sites $L$ and
with the boundary conditions $c_{0,n}=c_{1,n}$, $c_{L+1,n}=c_{L,n}$,
for any $n$.
The size of the lattice $L$ as well as the cut-off
number of atoms $N$ at each site were chosen large enough such that
their influences on the eigenstates are negligible~\cite{remark}.

\begin{figure}[t]
\begin{center}
\includegraphics[width=8cm]{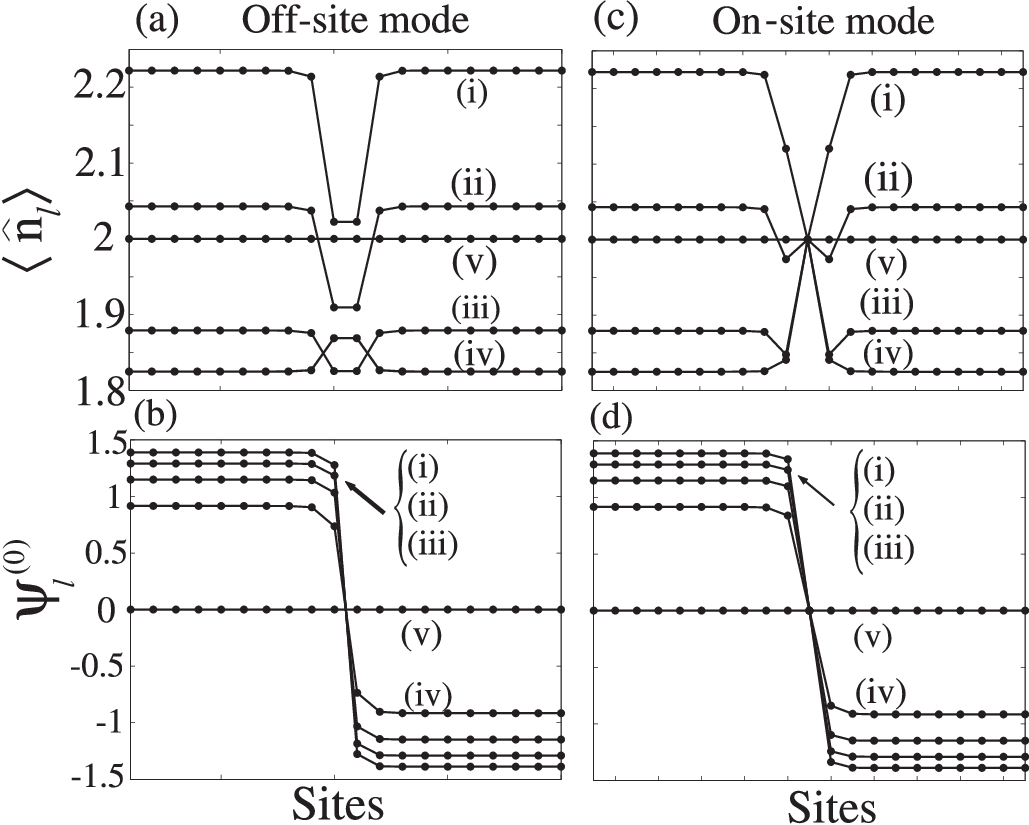}
\caption{
Mean number of atoms $\langle\hat{n}_l\rangle$ (a) and
(c), and mean-field order parameter $\psi_l^{(0)}$ (b) and (d).
The scaled chemical potential $\mu/U=1.2$ and the tunneling rates
$2dJ/U$: 0.7~(i), 0.5~(ii), 0.3~(iii), 0.15~(iv), and 0.05~(v).}
\label{smodes}
\end{center}
\end{figure}

Typical behavior of the modes is displayed in Fig.~\ref{smodes}.
The mean occupation numbers
$\langle\hat{n}_l\rangle$ calculated according to Eq.~(\ref{n})
are shown in (a) and (c), while (b) and (d) give the associated $\psi_l^{(0)}$
defined by Eq.~(\ref{psi}).
The individual curves correspond to different tunneling rates $J$.
Far from the middle point of the lattice,
$\langle\hat{n}_l\rangle$ as well as $\psi_l^{(0)}$ tend to the same values as
in the ground state. Near the middle point, on the other hand, they have nontrivial position dependence.

For the considered chemical potential, $\mu/U=1.2$, the MI-SF transition occurs at
$2dJ_c/U\approx0.0727$. Much above this value,
$\langle\hat{n}_l\rangle$ has only one extremum which is a global minimum.
It is doubly degenerate in the case of the off-site modes
[Fig.~\ref{smodes}(a), curves (i)-(iii); Fig.~\ref{smodes}(c), curve (i)].
Expectedly, these solutions reproduce the well-known standing soliton of the
DNSE~\cite{lattsol1}. For smaller values of $J$, when we come closer
to the phase boundary, the global minimum turns into a maximum
[Fig.~\ref{smodes}(a), curve (iv); Fig.~\ref{smodes}(c), curves (ii)-(iv)].
For the off-site modes, this maximum is always a global extremum.
In the case of the on-site modes, the maximum of $\langle\hat{n}_l\rangle$
is either a global extremum [Fig.~\ref{smodes}(c), curve (iv)]
or a local one which is accompanied by side minima
[Fig.~\ref{smodes}(c), curves (ii),~(iii)].
Contrary to the results deep in the SF region, these types of the atomic distributions cannot be described by the DNSE.

%----------------------------------------
%----------------------------------------

\begin{figure}[t]
\centerline{\includegraphics[width=8cm]{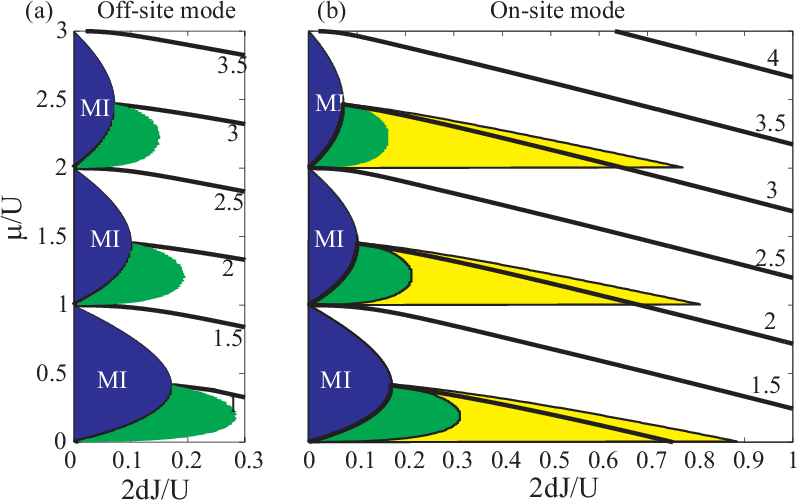}}
\caption{(Color online)
(a) {\it Off-site modes}.
Dark (blue) areas show the first three MI zones ($n_0=1,2,3$).
Grey (green) areas indicate the regions where the
off-site modes have a global maximum of $\langle\hat n_l\rangle$ at
the middle lattice sites around $l_0$, see curve (iv) of Fig.~\ref{smodes} (a).
In the rest part of the diagram, $\langle\hat n_l\rangle$ has only one extremum
and takes the minimal value at the middle sites,
see curves (i),~(ii),~(iii) of Fig.~\ref{smodes}~(a).
The lines of constant $\langle\hat{n}\rangle$ corresponding to the ground-state
densities are shown as well and labeled by the numerical values.\\
(b) {\it On-site modes}.
Grey (green) areas depict the regions where the on-site modes have a global maximum of
$\langle\hat n_l\rangle$ at the middle lattice site $l_0$,
(in these regions $\langle\hat n_l\rangle$ have only one extremum
which is a global maximum), see curve (iv) of Fig.~\ref{smodes} (c).
In the light-grey (yellow) areas,
the on-site modes have side minima near the maximum of $\langle\hat n_l\rangle$,
see curves (ii) and (iii) of Fig.~\ref{smodes} (c).
In the rest part of the diagram, $\langle\hat n_l\rangle$ has only one extremum
and takes the minimal value at the middle site,
see curve (i) of Fig.~\ref{smodes}~(c).
}
\label{fig2}
\end{figure}

In order to have a better understanding of the modes with the maxima of
$\langle\hat{n}_l\rangle$, we depict in Fig.~\ref{fig2}
a $(\mu,J)$-diagram identifying the various types of solutions.
The anomalous regions where $\langle\hat{n}_l\rangle$ attains a global
maximum are almost the same for the off-site and on-site modes.
They are, to a very good approximation, located in the ``hole"-areas of the
$(\mu,J)$-plane as displayed in Fig.~\ref{hs},
and hence, the corresponding modes can be interpreted as dark solitons of holes.
The anomalous regions of the on-site modes which have minima of
$\langle\hat{n}_l\rangle$ near the middle lattice point are located
in the intermediate regions between particle and hole-areas
and can thereby be interpreted as a mixture of dark solitons of holes and particles.
With the increase of the filling factor the size of the MI lobes as well as of
the anomalous regions decrease.

%----------------------------------------
%----------------------------------------

In the MI phase, the numerical procedure described above would give $\psi_l^{(0)}\equiv 0$.
This is because in the ground state $c_n^{(0)}=\delta_{n,n_0}$, i.e. $\psi^{(0)}=0$.
In this case there is no coupling between different lattice sites
in Eqs.~(\ref{GEd}) and the initial MI state remains unchanged during the imaginary-time
evolution. In order to introduce coupling between the lattice sites in Eqs.~(\ref{GEd}),
we have taken initial conditions for the same value of $\mu$ but for larger value
of $J$ corresponding to the SF phase. Nevertheless, the imaginary-time evolution
leads to the trivial result $\psi_l^{(0)}\equiv 0$.
The impossibility to get soliton solutions in the MI phase follows from the fact that
in the Gutzwiller ansatz, the excited states of the MI are products of
local Fock states, where the occupation numbers $n_l$ can be locally different from
the homogeneous filling $n_0$. As a consequence, since the system state is a product of Fock states at different sites, all $\psi_l^{(0)}$ must identically vanish
and no soliton solutions are therefore possible within these parameter regimes.

%----------------------------------------
\section{Stability of standing solitons}
%----------------------------------------

In this section we study the stability of the standing solitons with respect
to small perturbations.

%----------------------------------------
\subsection{Linear stability}
%----------------------------------------

We consider small perturbation of the soliton state determined by the coefficients
$c_{ln}^{(0)}$ as follows
$
c_{ln}(t)
=
\left[
    c_{ln}^{(0)}
    +
    c_{ln}^{(1)}(t)
\right]
\exp
\left(
    -i \omega_l t
\right)
$,
where $\omega_l$ is given by Eq.~(\ref{omegal}) and
\begin{equation}
\label{sol}
c_{ln}^{(1)}(t)
=
u_{ln}
e^{-i \omega t}
+
v_{ln}^*
e^{i \omega t}
\;.
\end{equation}
Substituting this expression into the Gutzwiller equations and keeping only linear terms
with respect to $u_{ln}$ and $v_{ln}$, we obtain the system of linear equations
\begin{eqnarray}
\label{evpexc}
\hbar\omega
u_{ln}
&=&
\sum_{n',l'}
\left(
    A_{nl}^{n'l'}
    u_{l'n'}
    +
    B_{nl}^{n'l'}
    v_{l'n'}
\right)
\;,
\nonumber\\
-
\hbar\omega
v_{ln}
&=&
\sum_{n',l'}
\left(
    B_{nl}^{n'l'}
    u_{l'n'}
    +
    A_{nl}^{n'l'}
    v_{l'n'}
\right)
\;,
\end{eqnarray}
where
\begin{eqnarray}
A_{nl}^{n'l'}
&=&
\left[
    \frac{U}{2}\,n(n-1)
    -
    \mu n
    -
    \hbar\omega_{l}
\right]
\delta_{l',l}
\delta_{n',n}
\nonumber\\
&-&
J
\left[
    \psi_{l-1}^{(0)}
    +
    \psi_{l+1}^{(0)}
    +
    2(d-1)
    \psi_{l}^{(0)}
\right]
\delta_{l',l}
\nonumber\\
&\times&
\left(
    \sqrt{n'}\,
    \delta_{n',n+1}
    +
    \sqrt{n}\,
    \delta_{n,n'+1}
\right)
\nonumber\\
&-&
J
\left[
    \sqrt{n+1}\,
    \sqrt{n'+1}\,
    c_{l,n+1}^{(0)}\,
    c_{l',n'+1}^{(0)}
\right.
\nonumber\\
&+&
\left.
\sqrt{n}\,
\sqrt{n'}\,
c_{l,n-1}^{(0)}\,
c_{l',n'-1}^{(0)}
\right]
\nonumber\\
&\times&
\left[
    \delta_{l,l'+1}
    +
    \delta_{l',l+1}
    +
    2(d-1)
    \delta_{l',l}
\right]
\;,
\nonumber\\
B_{nl}^{n'l'}
&=&
-J
\left[
    \sqrt{n+1}\,
    \sqrt{n'}\,
    c_{l,n+1}^{(0)}\,
    c_{l',n'-1}^{(0)}
\right.
\nonumber\\
&+&
\left.
    \sqrt{n}\,
    \sqrt{n'+1}\,
    c_{l,n-1}^{(0)}\,
    c_{l',n'+1}^{(0)}
\right]
\nonumber\\
&\times&
\left[
    \delta_{l,l'+1}
    +
    \delta_{l',l+1}
    +
    2(d-1)
    \delta_{l',l}
\right]
\;.
\nonumber
\end{eqnarray}
Eqs.~(\ref{evpexc}) are analogous to the Bogoliubov-de~Gennes equations which were employed
for the stability analysis of the dark solitons governed by the DNSE~\cite{JK99,KCTFM}.

\begin{figure}[t]
\begin{center}

%\psfrag{r}[c]{$\hbar\omega_r/U$}
%\psfrag{i}[c]{$\hbar\omega_i/U$}

%\includegraphics[width=6cm]{exc_ds_d_3-J_0.15-mu_1.2.eps}

%\includegraphics[width=6cm]{exc_ds_d_3-J_0.15-mu_1.4.eps}

%\includegraphics[width=6cm]{exc_ds_d_3-J_0.15-mu_2.eps}

\includegraphics[width=8cm]{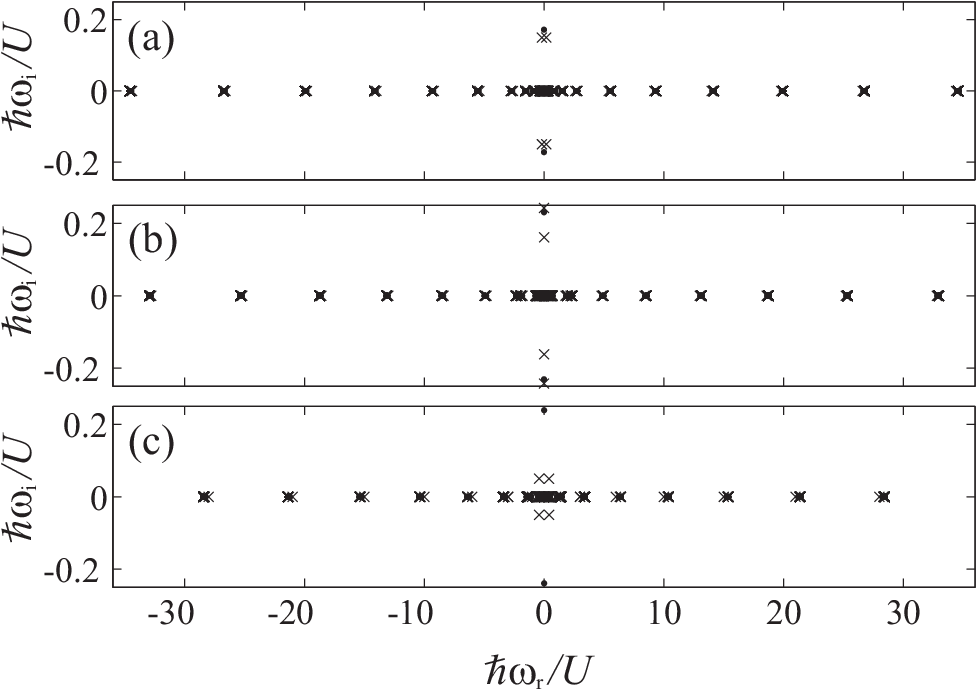}

\caption{
Spectrum of linear excitations of the off-site (dots) and on-site (crosses)
standing solitons for the tunneling rate $2dJ/U$=0.15.
Nonvanishing imaginary parts $\omega_i$ signal instability of the solitons.
The scaled chemical potential $\mu/U=1.2$~(a), $1.4$~(b), $2$~(c).}
\label{exc0.15}
\end{center}
\end{figure}

The stationary modes are linearly stable, if all the eigenvalues $\hbar\omega$
are real. Typical results of the solution of the eigenvalue problem (\ref{evpexc})
are shown in Fig.~\ref{exc0.15}. Most of the eigenvalues are real
but we get always few ones, which contain nonvanishing imaginary part $\omega_i$.
The magnitude of $\omega_i$ determines the inverse lifetime of the solitons,
which can be almost equal (see, e.g., Fig.~\ref{exc0.15}a)
or drastically different (like in Fig.~\ref{exc0.15}c) for the off-site and on-site modes.
We did not find any principal difference between the normal and anomalous modes.

%----------------------------------------
%----------------------------------------

\begin{figure}[t]
\begin{center}

%\psfrag{x}[c]{$\mu/U$}
%\psfrag{y}[c]{$2dJ/U$}
%\psfrag{z}[b]{max $\hbar\omega_i/U$}

%\includegraphics[width=7cm]{maxim_ds-d_3-L_50.eps}

%\includegraphics[width=7cm]{maxim_ds-d_3-L_51.eps}

\hspace{-2.5cm}
\includegraphics[width=9cm]{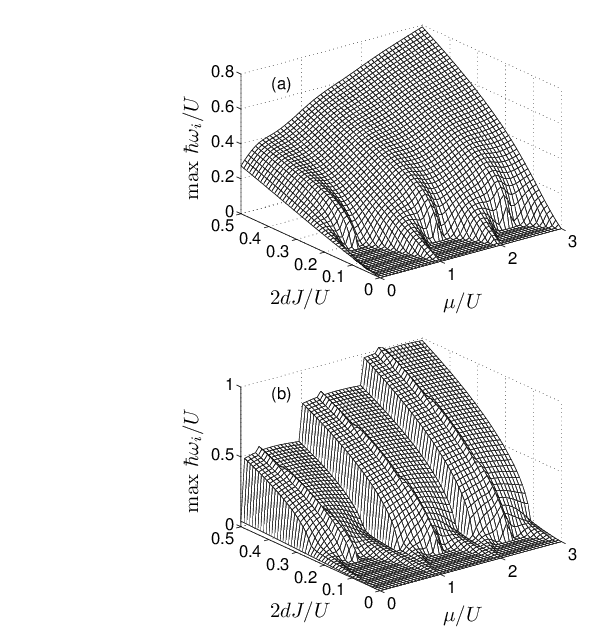}

\caption{
Maximal imaginary part of the complex eigenvalues $\omega$ for the
(a) off-site and (b) on-site modes.}
\label{maxim}
\end{center}
\end{figure}

Fig.~\ref{maxim} shows the maximal imaginary part of the complex eigenvalues $\omega$
which vanishes in the MI regions, where solitons do not exist, but does not vanish in the SF region.
With the increase of $\mu$ and $J$, maximal $\omega_i$ increases for both types of soliton modes meaning that the instability grows. There is, however, one important
qualitative difference between the off-site and on-site modes. For the on-site modes,
there are rather small regions between the MI lobes, where $\omega_i$ is close to zero and
much smaller than that for the off-site modes, i.e., the on-site solitons are much more stable.
This feature has some similarity to the stability of the standing dark solitons governed by the DNSE,
where it was found~\cite{JK99} that on-site modes are stable if the tunneling $J$
does not exceed a certain critical value, while off-site modes are unstable for all tunnelings.

%----------------------------------------
%----------------------------------------

%----------------------------------------
\subsection{Modulational stability}
%----------------------------------------

In order to study modulational stability of the standing solitons,
we perturb the eigenmodes calculated in the previous section by means
of transformation
\begin{equation}
c_{ln}
\to
{\cal N}_l
c_{ln}
\left(
    1 + \varepsilon_{ln}
\right)
\;,
\end{equation}
where $\varepsilon_{ln}$ are random numbers uniformly distributed in the interval
$[-\Delta/2,\Delta/2]$ and ${\cal N}_l$ are normalization constants
which are to be introduced in order to satisfy the normalization conditions~(\ref{norm}).
After that we solve numerically Eqs.~(\ref{GEd}) in real time.

\begin{figure}[t]
%\begin{center}

%\psfrag{x}[c]{$\tau$}
%\psfrag{y}[b]{sites}

%\includegraphics[width=4cm]{nav-d_3-J_0.15-mu_1.2-L_200.eps}
%\includegraphics[width=4cm]{nav-d_3-J_0.15-mu_1.2-L_201.eps}

%\includegraphics[width=4cm]{psi2-d_3-J_0.15-mu_1.2-L_200.eps}
%\includegraphics[width=4cm]{psi2-d_3-J_0.15-mu_1.2-L_201.eps}

\hspace{-3.3cm}
\includegraphics[width=11cm]{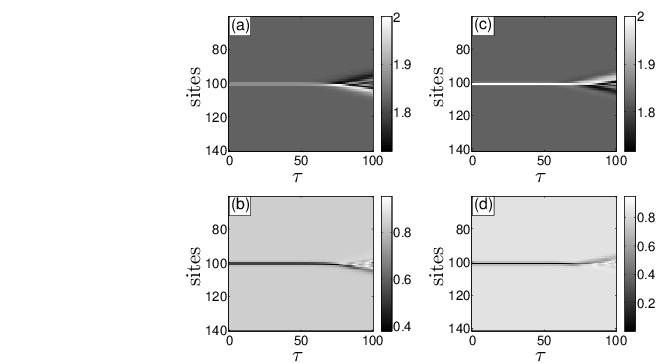}

\caption{
Time evolution of the mean number of atoms $\langle\hat{n}_l\rangle$ (a,c)
and the condensate $\left|\psi_l\right|^2$ (b,d) due to the
slight perturbation of the off-site (a,b) as well as on-site (c,d)
standing solitons with $\Delta/U=10^{-4}$.
The parameters are $2dJ/U=0.15$ and $\mu/U=1.2$.
$\tau=tU/\hbar$ is the dimensionless time.}

\label{J_0.15-mu_1.2}
%\end{center}
\end{figure}

\begin{figure}[t]
\begin{center}

%\psfrag{x}[c]{$\tau$}
%\psfrag{y}[b]{sites}

%\includegraphics[width=4cm]{nav-d_3-J_0.15-mu_2-L_200.eps}
%\includegraphics[width=4cm]{nav-d_3-J_0.15-mu_2-L_201.eps}

%\includegraphics[width=4cm]{psi2-d_3-J_0.15-mu_2-L_200.eps}
%\includegraphics[width=4cm]{psi2-d_3-J_0.15-mu_2-L_201.eps}

%\hspace{-3.3cm}
%\includegraphics[width=11cm]{f7.eps}

\hspace{-3.3cm}
\includegraphics[width=11cm]{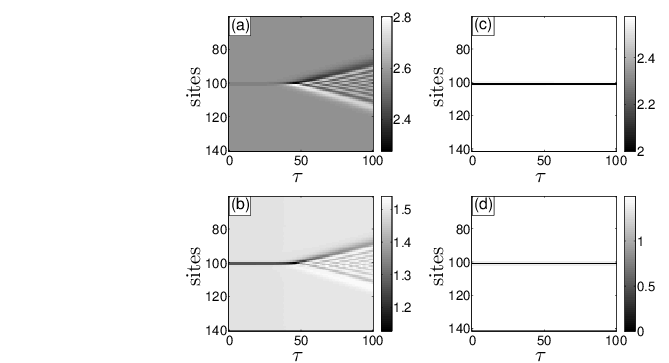}

\caption{
The same as in Fig.~\ref{J_0.15-mu_1.2} but for $\mu/U=2$.}

\label{J_0.15-mu_2}
\end{center}
\end{figure}

The results of these calculations are shown
in Figs.~\ref{J_0.15-mu_1.2},~\ref{J_0.15-mu_2} and they are closely connected
to the linear stability analysis of the previous subsection.
In the example shown in Fig.~\ref{J_0.15-mu_1.2}, the off-site and on-site solitons
remain stable within the same time interval because the imaginary parts of
the spectrum of linear excitations are approximately the same for the two modes
(see Fig.~\ref{exc0.15}a). On the other hand, Fig.~\ref{J_0.15-mu_2} shows an
example when the lifetime of the off-site mode is much shorter than that of
the on-site one which is consistent with the fact that the imaginary part of
the complex eigenvalue is larger for the off-site mode (see Fig.~\ref{exc0.15}c).

%----------------------------------------
%----------------------------------------

%----------------------------------------
%----------------------------------------

%----------------------------------------
\section{Propagating solitons created by phase imprinting}
%----------------------------------------

Experimentally, dark (or grey) solitons are typically created via a
phase-imprinting method~\cite{maciek-sengstock,denschlag}.
Initially ($t=0$) the system of atoms is assumed to be in its ground state.
During a short time $t_{imp}$ one applies a spatially dependent
potential on top of the lattice. In the Bose-Hubbard Hamiltonian,
it is described by the term $\sum_l \epsilon_l \hat a_l^\dagger \hat a_l$.
If the time $t_{imp}$ is much shorter than other characteristic time scales,
from Eqs.~(\ref{GEd}) we get that the additional term induces a shift in the
phase of the atomic states
\begin{eqnarray}
c_{ln}(t_{imp})
&=&
c_n^{(0)}
\exp
\left(
    - i \phi_l n
\right)
\;,
\nonumber\\
\psi_{l}(t_{imp})
&=&
\psi^{(0)}
\exp
\left(
    - i \phi_l
\right)
\;.
\end{eqnarray}
For the creation of dark solitons we choose a hyperbolic tangent
imprinting potential, such that
\begin{equation}
\phi_l
=
\frac{\epsilon_l t_{imp}}{\hbar}
=
\frac{\Delta\phi}{2}
\left[
    1+
    \tanh
    \left(
        \frac{l-l_0}{0.45\, l_{imp}}
    \right)
\right]
\;,
\end{equation}
where $l_0$ is the middle point of the lattice.
Here, $l_{imp}$ is the width of the interval around $l=l_0$ where
$\phi_l/\Delta\phi$ grows from $0.1$ to $0.9$, and
$\Delta\phi$ is the amplitude of the imprinted phase~\cite{imprintfunc}.
Apart from the moving grey soliton, the phase imprinting also
induces a density wave propagating in the opposite direction to the
soliton, which appears due to the impulse imparted by the imprinting potential~\cite{maciek-sengstock,denschlag,imprintfunc}.

%----------------------------------------
\begin{figure}[t]

%\psfrag{x}[c]{$\tau$}
%\psfrag{y}[b]{sites}

%\includegraphics[width=8cm]{nav-d_3-J_0.3-mu_1.2-w_2-L_201.eps}

%\includegraphics[width=8cm]{psi2-d_3-J_0.3-mu_1.2-w_2-L_201.eps}

\hspace{-3.3cm}
\includegraphics[width=11cm]{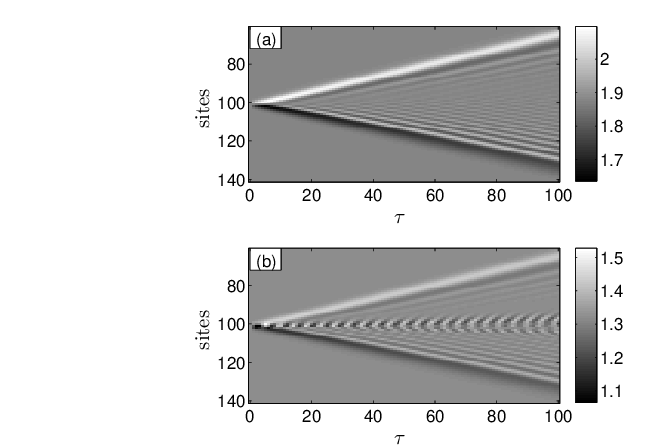}

\caption{
Time evolution of the mean occupation numbers $\langle\hat{n}_l\rangle$ (a)
and the order parameters $|\psi_l|^2$ (b)
after phase imprinting with $\Delta\phi=\pi$, $l_{imp}=2$.
The parameters are $\mu/U=1.2$, $2dJ/U=0.3$, giving similar evolution for both
$\langle\hat{n}_l\rangle$ and $|\psi_l|^2$.
$\tau=tU/\hbar$ is the dimensionless time.}
\label{fig3}
\end{figure}
%----------------------------------------

\begin{figure}[t]

%\psfrag{x}[c]{$\tau$}
%\psfrag{y}[b]{sites}

%\includegraphics[width=8cm]{nav-d_3-J_0.15-mu_1.2-w_2-L_201.eps}

%\includegraphics[width=8cm]{psi2-d_3-J_0.15-mu_1.2-w_2-L_201.eps}

\hspace{-3.3cm}
\includegraphics[width=11cm]{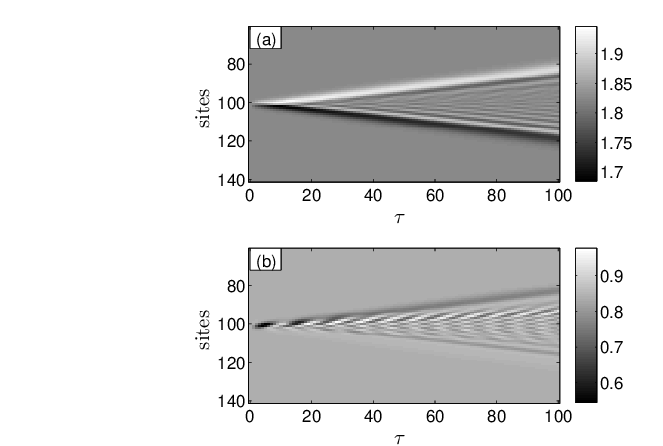}

\caption{
Same as in Fig.~\ref{fig3}, but with $2dJ/U=0.15$. In this regime,
the directions of propagating maximum and minimum of $|\psi_l|^2$ are
reversed compared to those of Fig.~\ref{fig3}.
$\tau$ is the dimensionless time.}
\label{fig4}
\end{figure}

\begin{figure}[t]

%\psfrag{x}[c]{$\tau$}
%\psfrag{y}[b]{sites}

%\includegraphics[width=8cm]{nav-d_3-J_0.15-mu_1.4-w_2-L_201.eps}

%\includegraphics[width=8cm]{psi2-d_3-J_0.15-mu_1.4-w_2-L_201.eps}

\hspace{-3.3cm}
\includegraphics[width=11cm]{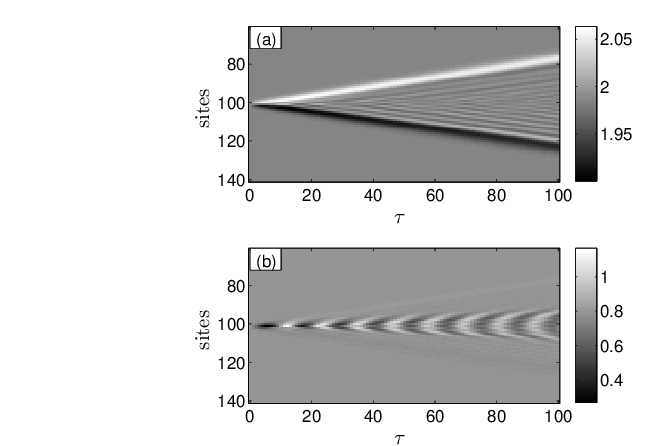}

\caption{
Same as in Figs.~\ref{fig3} and~\ref{fig4}, but with $\mu/U=1.4$, $2dJ/U=0.15$.
For these parameters, $|\psi_l|^2$ oscillates and spreads near the imprinting site.
$\tau$ is the dimensionless time.}
\label{fig5}
\end{figure}

Figs.~\ref{fig3},~\ref{fig4},~\ref{fig5} present the time evolution
of $\langle \hat{n}_l\rangle$ and $|\psi_l|^2$ for two different
values of $J$ and $\mu$, corresponding to different regions of the
diagram in Fig.~\ref{fig2}. Scaling $\mu$ and $J$ by $U$, we work
with a dimensionless time $\tau=tU/\hbar$. In Fig.~\ref{fig3}, $J$
is taken relatively large but still in the regime where the on-site
modes have local maxima of $\langle \hat{n}_l\rangle$ [light-grey (yellow)
region in Fig.~\ref{fig2}~(b)]. In this case, the dips and the maxima of
$\langle \hat{n}_l\rangle$ coincide with those of $|\psi_l|^2$ but
the dips propagate slower than the maxima. This type of dynamics is
always recovered in simulations based on the DNSE. Therefore, in
this regime the usual dark solitons can still be created by the
phase imprinting, although the standing on-site soliton modes are somewhat
distorted.

In Fig.~\ref{fig4}, the value of $\mu$ is the same as in
Fig.~\ref{fig3} but $J$ is smaller such that we enter into the grey (green)
region of Fig.~\ref{fig2}, where the standing solitons have global maxima
of $\langle\hat{n}_l\rangle$.
During the time evolution, the overall structure of $\langle\hat{n}_l\rangle$
remains as in the example shown in Fig.~\ref{fig3},
but here $|\psi_l|^2$ behaves differently.
More precisely, $|\psi_l|^2$ shows a local maximum coinciding with the
propagating dip of the occupation numbers, and a local minimum where
instead $\langle \hat{n}_l\rangle$ has a local maximum.
This anomalous behavior,
which is found only within the grey (green) regions of Fig.~\ref{fig2}
as a manifestation of the hole superfluidity,
cannot be described by the DNSE. Closer to the boundary of the grey (green)
region as in the example depicted in Fig.~\ref{fig5}, $|\psi_l|^2$
can become oscillating and spreading around the imprinting site,
while $\langle\hat{n}_l\rangle$ still shows similar dynamics as in
Figs.~\ref{fig3} and \ref{fig4}. In the remaining white regions of
Fig.~\ref{fig2}, the time evolution is qualitatively the same as in
Fig.~\ref{fig3}.
The simulations for $l_0$ on the lattice site and in the middle of two
neighboring sites do not show any noticeable difference in the time evolution.

Since the difference in the dynamics shows up in the behavior of $\psi_l$,
it should be possible to observe it in the time-of-flight
experiments~\cite{BDZ}
which are based on the measurement of the momentum distribution
\begin{equation}
P({\bf k})
=
\left|
    W({\bf k})
\right|^2
\sum_{{\bf i},{\bf j}}
\langle
    \hat a_{\bf i}^\dagger
    \hat a_{\bf j}
\rangle
\exp
\left[
    i {\bf k} \cdot
    \left(
        {\bf i} - {\bf j}
    \right)
\right]
\;,
\nonumber
\end{equation}
where $W({\bf k})$ is the Fourier transform of the Wannier function~\cite{BDZ}.
In the Gutzwiller approximation, it takes the form
\begin{eqnarray}
P({\bf k})
&=&
\left|
    W({\bf k})
\right|^2
\left[
    \sum_{\bf i}
    \left(
        \langle
             \hat n_{\bf i}
        \rangle
        -
        \left|
            \psi_{\bf i}
        \right|^2
    \right)
\right.
\nonumber\\
&&
\left.
    +
    \left|
    \sum_{\bf i}
    \psi_{\bf i}^*
    \exp
    \left(
        i {\bf k} \cdot {\bf i}
    \right)
    \right|^2
\right]
\;.
\nonumber
\end{eqnarray}
Due to the fact that we are dealing here with infinite lattices,
$P({\bf k})$ contains singular contributions which makes it difficult to compare
different regimes. Real experiments are always done in finite lattices
in the presence of the harmonic trap, where there are no singularities.
Therefore, in order to make more concrete experimental predictions,
further investigations beyond the scope of the present work are required.
{\it In situ} imaging of ultracold atoms in optical lattices became also possible
due to novel techniques~\cite{nl}
which allow to measure individual site occupations  $\langle\hat n_{\bf i}\rangle$ as well.

%----------------------------------------
%----------------------------------------

%----------------------------------------
%----------------------------------------

We also performed simulations with other values of the width
$l_{imp}$ and did not find any strong influence on the propagation
velocity. Larger values of $l_{imp}$ result in broader propagating
modes and their shapes become more smooth. However, we did find that
the interference pattern between the modes propagating in the
opposite directions, visible in the center of
Figs.~\ref{fig3},~\ref{fig4},~\ref{fig5}, becomes suppressed for
larger $l_{imp}$.

\begin{figure}[t]

\centerline{\includegraphics[width=7cm]{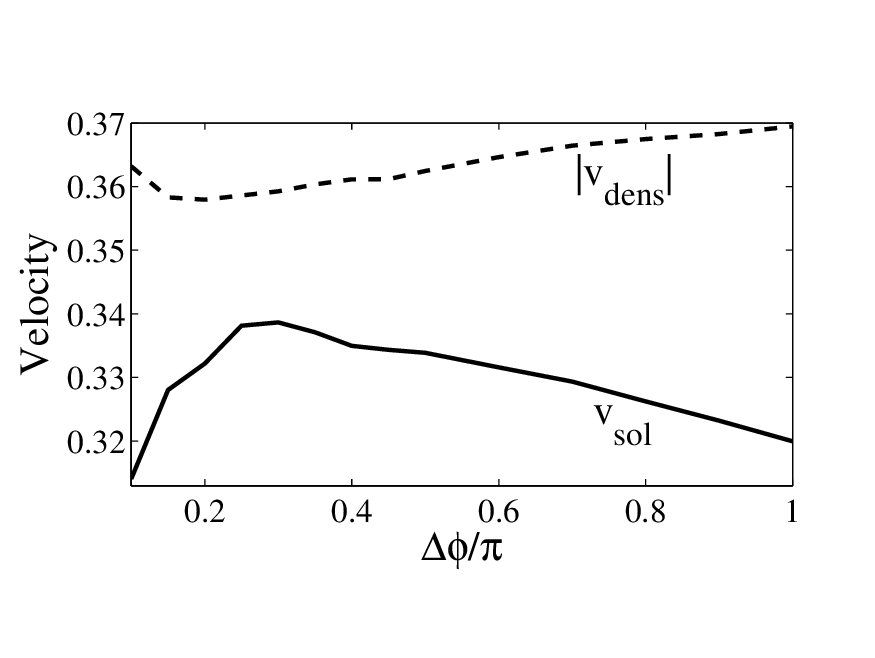}}

\vspace{-1.1cm}

\caption{Dimensionless velocity of the soliton (solid line) and
density (dashed line) modes as a function of $\Delta\phi$.
Here $2dJ/U=0.3$, $\mu/U=1.2$, and $l_{imp}=2$.}
\label{fig6}
\end{figure}

Finally, the soliton velocity $v_{sol}$ as well as the velocity of
the density wave $v_{dens}$ as functions of $\Delta\phi$ are shown
in Fig.~\ref{fig6}. They are calculated making use of a linear fit
of the corresponding minimum and maximum of $\langle \hat n_l\rangle$ for
the dimensionless time $\tau=tU/\hbar>10$. Although the dependences
are very weak, it is found that the velocities are not monotonic
showing a maximum for $\Delta\phi\approx\pi/4$. This behavior is
different from what was found in the simulations based on the
Gross-Pitaevskii equation in continuum in a harmonic
trap~\cite{maciek-sengstock} or in a periodic
potential~\cite{yulin}. On the other hand, for increasing soliton velocity, the depth of the propagating dip as well as the height
of its maxima decreases. The same was found in continuum
model~\cite{maciek-sengstock}.
However, in the continuum model, where analytical expressions for moving grey soliton solutions are known, it directly follows that in proper units the square of the soliton depth plus the square of the soliton velocity is a constant, i.e. increasing the velocity makes the soliton more shallow~\cite{maciek-sengstock}. The present analysis differs in several aspects from the continuum one; the solitons are propagating within a lattice, we consider dynamics beyond the regular mean-field, and the studied moving solitons are not time-dependent solutions of the system Hamiltonian but rather phase imprinted soliton-like states. Thereby, not surprisingly, we have numerically found that the same velocity-depth relation does not hold in our situation. In particular, our results indicate that for small-to-moderate imprinting amplitudes $\Delta\phi$ (as in Fig.~\ref{fig6}) the velocity-depth relation fails, while for increasing amplitudes it holds better.

%----------------------------------------
%----------------------------------------

%----------------------------------------
\section{Conclusion}
%----------------------------------------

We have investigated dark solitons of bosons in
optical lattices at zero temperature. Using the Gutzwiller ansatz,
we found kink stationary states where the condensate order parameter
$\psi_l$ is an antisymmetric function with respect to some spatial
point. However, in certain regions close to the MI-SF transition
which are shown in Fig.~\ref{fig2}, the corresponding
$\langle\hat n_l\rangle$ does not have a global minimum at the point where
$\psi_l$ vanishes. This anomalous behavior shows up near the phase
boundary where one has to distinguish between particle and hole superfluidity.
In general, we have found three types of stationary states:
solitons of particles, solitons of holes and a mixture of both.
A stability analysis revealed that the soliton solutions are sensitive to small perturbations
and, therefore, unstable. Their lifetime differs in general for the on-site and off-site modes.
For the on-site modes, there are small regions between the MI lobes,
where the lifetime is very large and much larger than that for the off-site modes.

The real-time dynamics of the propagating dark solitons created by
the phase imprinting is studied as well. In the MI phase, the
solitons cannot be created. This can be done only in the SF phase,
where there are always global minima and maxima of $\langle\hat n_l\rangle$ propagating
in the opposite directions and these directions are always the same.
The behavior of the condensate $\left|\psi_l\right|^2$ can be qualitatively different
and this happens in the anomalous regions of the stationary modes,
where $\langle\hat n_l\rangle$
has a global maximum as a manifestation of the hole superfluidity.

%----------------------------------------

\begin{acknowledgments}
We would like to thank R.~Graham and
B.~Malomed for helpful discussions. The work of KK was supported by
the SFB/TR 12 of the German Research Foundation (DFG). JL
acknowledges support from VR-Vetenskapsr\aa det and the MEC program
(FIS2005-04627). ML acknowledges Spanish MEC/MINCIN projects TOQATA
(FIS2008-00784) and QOIT (Consolider Ingenio 2010), ESF/MEC project
FERMIX (FIS2007-29996-E), EU STREP project NAMEQUAM, ERC Advanced
Grant QUAGATUA, and Alexander von Humboldt Foundation Senior
Research Prize.
\end{acknowledgments}

%----------------------------------------


\begin{thebibliography}{999}
%----------------------------------------

\bibitem{stringari}
L.~P.~Pitaevskii and S.~Stringari,
{\it Bose-Einstein Condensation},
Oxford University Press, Oxford, 2003.

\bibitem{kivshar}
Y.~S.~Kivshar and G.~P.~Agrawal,
{\it Optical Solitons: From Fibers to Photonic Crystals},
Academic Press, 2003.

\bibitem{maciek-sengstock}
S.~Burger, K.~Bongs, D.~Dettmer, W.~Ertmer, K.~Sengstock, A.~Sanpera, G.~V.~Shlyapnikov,
and M.~Lewenstein,
Phys.~Rev.~Lett. {\bf 83}, 5198 (1999).

\bibitem{denschlag}
J.~Denschlag, J.~E.~Simsarian, D.~L.~Feder, C.~W.~Clar, L.~A.~Collins, J.~Cubizolles, L.~Deng, E.~W.~Hagley, K.~Helmerson, W.~P.~Reinhardt, S.~L.~Rolston, B.~I.~Schneider,
and W.~D.~Phillips,
Science {\bf 287}, 97 (2000).

\bibitem{Stellmer08}
S.~Stellmer, C.~Becker, P.~Soltan-Panahi, E.-M.~Richter, S.~D\"orscher,
M.~Baumert, J.~Kronj\"ager, K.~Bongs, and K.~Sengstock,
Phys.~Rev.~Lett. {\bf 101}, 120406 (2008).

\bibitem{Weller08}
A.~Weller, J.~P.~Ronzheimer, C.~Gross, J.~Esteve, M.~K.~Oberthaler,
D.~J.~Frantzeskakis, G.~Theocharis, and P.~G.~Kevrekidis,
Phys.~Rev.~Lett. {\bf 101}, 130401 (2008).

\bibitem{Theocharis}
G.~Theocharis, A.~Weller, J.~P.~Ronzheimer, C.~Gross, M.~K.~Oberthaler, P.~G.~Kevrekidis,
and D.~J.~Frantzeskakis,
arXiv:0909.2122.

\bibitem{khaykovich/strecker}
L.~Khaykovich, F.~Schreck, G.~Ferrari, T.~Bourde, J.~Cubizolles, L.~D.~Carr, Y.~Castin,
and C.~Salomon,
Science {\bf 296}, 1290 (2002);
K.~E.~Strecker, G.~B.~Partridge, A.~G.~Truscott, and R.~G.~Hulet,
Nature {\bf 417}, 150 (2002).

\bibitem{sengstock}
C.~Becker, S.~Stellmer, P.~Soltan-Panahi, S.~D\"orscher, M.~Baumert, E.-M.~Richter,
J.~Kronj\"oger, K.~Bongs, and K.~Sengstock,
Nature Physics {\bf 4}, 496 (2008).

\bibitem{Shomroni09}
I.~Shomroni, E.~Lahoud, S.~Levy, and J.~Steinhauer,
Nature Physics {\bf 5}, 193 (2009).

\bibitem{smerzi}
A.~Trombettoni and A.~Smerzi,
Phys.~Rev.~Lett. {\bf 86}, 2353 (2001).

\bibitem{anna}
V.~Ahufinger and A.~Sanpera,
Phys.~Rev.~Lett. {\bf 94}, 130403 (2005).

%\bibitem{gapexp}
\bibitem{oberthaler}
B.~Eiermann, Th.~Anker, M.~Albiez, M.~Taglieber, P.~Treutlein, K.-P.~Marzlin,
and M.~K.~Oberthaler,
Phys.~Rev.~Lett. {\bf 92}, 230401 (2004).
% Experimental gap solitons with BEC %%%

\bibitem{muryshev}
A. Muryshev, G. V. Shlyapnikov, W. Ertmer, K. Sengstock, and M. Lewenstein,
Phys.~Rev.~Lett. {\bf 89}, 110401 (2002).
%B.~Jackson {\it et al.} Phys.~Rev.~A {\bf 75}, 051601(R) (2007).

\bibitem{MR}
A.~D.~Martin and J.~Ruostekoski,
Phys.~Rev.~Lett. {\bf 104}, 194102 (2010).

\bibitem{dziarmaga}
J.~Dziarmaga, Phys.~Rev.~A {\bf 70}, 063616 (2004).
%J.~Dziarmaga and K.~Sacha, Phys.~Rev.~A {\bf 66}, 043620 (2002).

\bibitem{yulin}
A.~V.~Yulin and D.~V.~Skryabin,
Phys.~Rev.~A {\bf 67}, 023611 (2003).

\bibitem{JK99}
M.~Johansson and Y.~S.~Kivshar,
Phys.~Rev.~Lett. {\bf 82}, 85 (1999).

\bibitem{KCTFM}
P.~G.~Kevrekidis, R.~Carretero-Gonz\'alez, G.~Theocharis, D.~J.~Frantzeskakis,
and B.~A.~Malomed,
Phys.~Rev.~A {\bf 68}, 035602 (2003).

\bibitem{castin}
Y.~Castin,
Eur.~Phys.~J.~B {\bf 68}, 317 (2009).

\bibitem{boris}
M.~Lewenstein and B.~A.~Malomed,
New~J.~Phys. {\bf 11}, 113014 (2009).

\bibitem{juha}
J.~Javanainen and U.~Shrestha,
Phys.~Rev.~Lett. {\bf 101}, 170405 (2008).

\bibitem{vidal}
G.~Vidal,
Phys.~Rev.~Lett. {\bf 93}, 040502 (2004).

\bibitem{carr}
R.~V.~Mishmash and L.~D.~Carr,
Math. Comput. Simul. {\bf 80}, 732 (2009) ;
R.~V.~Mishmash and L.~D.~Carr,
Phys.~Rev.~Lett. {\bf 103}, 140403 (2009);
R.~V.~Mishmash, I.~Danshita, Ch.~W.~Clark and L.~D.~Carr,
Phys.~Rev.~A {\bf 80}, 053612 (2009).

\bibitem{Fisher}
M.~P.~A.~Fisher, P. B. Weichman, G. Grinstein, and D. S. Fisher,
Phys.~Rev.~B {\bf 40}, 546 (1989).

\bibitem{WCHZ}
C.~Wu, H.~Chen, J.~Hu, and S.-C.~Zhang,
Phys.~Rev.~A {\bf 69}, 043609 (2004).

\bibitem{JBCGZ}
D.~Jaksch, C.~Bruder, J.~I.~Cirac, C.~W.~Gardiner, and P.~Zoller,
Phys.~Rev.~Lett. {\bf 81}, 3108 (1998).

%\bibitem{gaptheo}
%N. K. Efremidis and D. N. Christodoulides,
%Phys. Rev. A {\bf 67}, 063608 (2003);
%D. E. Pelinovsky {\it et al.},%, A. A. Sukhorukov, and Y. S. Kivshar,
%Phys. Rev. E {\bf 70}, 036618 (2004). %% Gap soliton theory %%%

\bibitem{BPVB07}
P.~Buonsante, V.~Penna, A.~Vezzani, P.~B.~Blakie,
Phys.~Rev.~A {\bf 76}, 011602(R) (2007).

\bibitem{Z}
J.~Zakrzewski,
Phys.~Rev.~A {\bf 71}, 043601 (2005).

\bibitem{GM}
E.~Lundh, Europhys.~Lett. {\bf 84}, 10007 (2009);
D.~S.~Goldbaum and E.~J.~Mueller,
Phys.~Rev.~A {\bf 79}, 021602(R) (2009).

\bibitem{phas}
R.~Balakrishnan, I.~I.~Satija, and C.~W.~Clark,
Phys.~Rev.~Lett. {\bf 103}, 230403 (2009).

\bibitem{Perez-Garcia}
V. M. P\'erez-Garc\'ia, H. Michinel, J. I. Cirac, M. Lewenstein, and P. Zoller,
Phys.~Rev.~Lett. {\bf 77}, 5320 (1996). 

\bibitem{SKPR}
K.~Sheshadri, H.~R.~Krishnamurthy, R.~Pandit and T.~V.~Ramakrishnan,
Europhys.~Lett. {\bf 22}, 257 (1993).

\bibitem{OSS}
D.~van~Oosten, P.~van~der~Straten, and H.~T.~Stoof,
Phys.~Rev.~A {\bf 63}, 053601 (2001).

\bibitem{Sachdev}
S.~Sachdev,
{\it Quantum Phase Transitions}
(Cambridge University Press, Cambridge, England, 2001).

\bibitem{Dalfovo}
F.~Dalfovo, S. Stringari, Phys.~Rev.~A {\bf 53}, 2477 (1996).

\bibitem{lattsol1}
Y.~S.~Kivshar, W.~Krolikowski, and O.~A.~Chubykalo,
Phys.~Rev.~E {\bf 50}, 5020 (1994); V. Ahufinger, A. Sanpera, P. Pedri, L. Santos, and M. Lewenstein, Phys. Rev. A {\bf 69}, 053604  (2004).

\bibitem{remark}
All the calculations presented in the figures are performed for $d=3$ and we used $N=10$.
In the calculations of the off-site and on-site modes, the lattice size was
$L=200$ and $L=201$, respectively, such that the middle lattice point was
$l_0=(L+1)/2$.

\bibitem{imprintfunc}
S.~Burger, L.~D.~Carr, P.~\"Ohberg, K.~Sengstock, and A.~Sanpera,
Phys.~Rev.~A  {\bf 65}, 043611 (2002).

\bibitem{Lew03}
B.~Damski, J.~Zakrzewski, L.~Santos, P.~Zoller, and M.~Lewenstein,
Phys.~Rev.~Lett. {\bf 91}, 080403 (2003).

%\bibitem{FM}
%J.~K.~Freericks and H.~Monien,
%Europhys.~Lett. {\bf 26}, 545 (1994);
%Phys.~Rev.~B {\bf 53}, 2691 (1996).

%\bibitem{ASPSL}
%V.~Ahufinger {\it et al.},%, A.~Sanpera, P.~Pedri, L.~Santos, and M.~Lewenstein,
%Phys.~Rev.~A {\bf 69}, 053604 (2004).

\bibitem{BDZ}
I.~Bloch, J.~Dalibard, and W.~Zwerger,
Rev.~Mod.~Phys. {\bf 80}, 885 (2008)
and references therein.

%\bibitem{GWRLO}
\bibitem{nl}
T.~Gericke, P.~W\"urtz, D.~Reitz, T.~Langen, and H.~Ott,
Nature Physics {\bf 4}, 949 (2008);
%
%\bibitem{GCHC}
N.~Gemelke, X.~Zhang, Ch.-L.~Hung, and Ch.~Chin,
Nature {\bf 460}, 995 (2009);
%
W.~S.~Bakr, J.~I.~Gillen, A.~Peng, S.~F\"olling, and M.~Greiner,
Nature {\bf 462}, 74 (2009);
%
%\bibitem{IVLMGS}
A.~Itah, H.~Veksler, O.~Lahav, A.~Blumkin, C.~Moreno, C.~Gordon, and J.~Steinhauer,
Phys.~Rev.~Lett. {\bf 104}, 113001 (2010);
%
W.~S.~Bakr, J.~I.~Gillen, A.~Peng, S.~F\"olling, and M.~Greiner,
arXiv:1006.0754.

\end{thebibliography}
\end{document}